\documentclass[review]{elsarticle}

\journal{Journal of \LaTeX\ Templates}

\usepackage{lineno,hyperref}
\modulolinenumbers[5]
\usepackage{amssymb}
\usepackage{amsmath}
\usepackage{epsfig}
\usepackage{amsthm}
\usepackage{tabularx}
\usepackage{graphicx}
\usepackage{caption}
\usepackage{mathtools}
\usepackage{enumerate}
\usepackage{color}
\usepackage{url}
\newcommand{\bfs}{\boldsymbol{\sigma}}

\newcommand{\bfj}{{\bf J}}

\newcommand{\Z}{{\mathbb Z}}

\newcommand{\reff}[1]{(\ref{#1})}

\newtheorem {lemma}{Lemma}

\newtheorem {theorem}{Theorem}
\newtheorem{rem}{Remark}

\begin{document}

\begin{frontmatter}

\title{Stochastic Ising model with plastic interactions}



\author[mymainaddress,mysecondaryaddress]{Eugene Pechersky}
\author[mymainaddress]{Guillem Via}
\author[mymainaddress]{Anatoly Yambartsev}



\address[mymainaddress]{Institute of Mathematics and Statistics, University of S\~ao Paulo, Brazil.}
\address[mysecondaryaddress]{Institute for Information Transmission Problem, Russian Academy of Science, Russia.}

\begin{abstract}
We propose a new model based on the Ising model with the aim to study synaptic plasticity phenomena in neural networks. It is today well established in biology that the synapses or connections between certain types of neurons are strengthened when the neurons are co-active, a form of the so called synaptic plasticity. Such mechanism is believed to mediate the formation and maintenance of memories.
The proposed model describes some features from that phenomenon. 
Together with the spin-flip dynamics, in our model the coupling constants are also subject to stochastic dynamics, so that they interact with each other. The evolution of the system is described by a continuous-time Markov jump process.
\end{abstract}


\end{frontmatter}


\section{Introduction}

To understand the mechanisms underlying the formation, maintenance and recall of memories in our brains is one of the main aims in neuroscience. In his pioneering work~\cite{hebb}, Hebb proposed the cell assembly hypotheses to explain them, triggering a plethora of theoretical and experimental studies that tried to test it \cite{historySTDP, schultscortexassembly, reviewattractors}. According to that hypotheses, memories are encoded in engrams or patterns of connectivity within certain subpopulations (or assemblies) of neurons. Each engram corresponds to a configuration of strengths for the connections or synapses between the neurons in the assembly \cite{reviewattractors}. Recall that neurons communicate with each other through short electric pulses or spikes. In particular, spikes emitted by a neuron bring closer to fire other neurons with which it is connected by synapses. In this way, neurons in an assembly activate each other and sustain the activity of the network, among other mechanisms, through the recurrent connections between them \cite{wangreverberatory}, which form the engram. It is actually been shown in theoretical studies how an engram can maintain several different activity patterns within the assembly that are stable for timescales much longer than those for the single neuron dynamics. Only one of these stable patterns of activity, also called attractors of the network, would be active at a time and would encode an object to be kept active in memory~\cite{reviewattractors}. The patterns can be defined as a configuration of firing rates (spikes per second) for each neuron or as a particular spatio-temporal sequence of spikes. In this framework the process of learning consists on the formation of the engram that will support a given set of attractors. In real biological systems that formation could happen through synaptic plasticity mechanisms~\cite{historySTDP, gerstnerkistler}, i.e. changes in the strengths of the synapses. Hebb \cite{hebb, historySTDP} also proposed that the strengthening of the synapses between co-active neurons could be responsible for that formation. This mechanism is known today as Hebbian plasticity and is a form of Long-Term Potentiation (LTP) and of activity-dependent plasticity. There is some experimental evidence supporting the idea that attractor states of cell assemblies encode memories in different regions of the brain~ \cite{reviewattractors, hippoattractor2}, and that the engrams supporting these attractors are formed by means of LTP~\cite{martingrimwood, nevescooke}. Moreover, some studies show the need of other forms of plasticity, like homeostatic plasticity~\cite{wangreverberatory, gerstnerkistler, homeostasisreview}, where synaptic changes are not triggered by neural activity but by internal biological mechanisms, in order to yield a stable formation and maintenance of engrams and memories. Consolidation mechanisms have also been shown to be necessary for such stability~\cite{gerstnerkistler, consolidation}.

Hopfield \cite{hopfield, amit} proposed a model to study the dynamics of attractors and the storage capacity of neural networks by means of the Ising model \cite{ising, Georgii}. In this framework each neuron would be represented by a spin whose up and down states correspond to a high and a low firing rates, respectively. Then the cell assembly would be represented by the set of vertices in the network, the engram by the connectivity matrix and the attractor by the stable spin configurations. Hopfield gave a mathematical expression for the connectivity matrix that supports a given set of attractors chosen \emph{a priori}. However, the learning phase in which such connectivity is built through synaptic plasticity mechanisms is not been considered within its framework. There exist some more biologically realistic models of neural networks which model the formation of engrams through synaptic plasticity mechanisms but the analysis was done only numerically and there is no analytical result in them \cite{reviewspikinglearning}. To the best of our knowledge, analytical results on neural networks with plastic synapses where the learning phase is been considered are restricted to models of binary neurons with binary synapses. In these studies, the convergence and other properties of the learning process, were proved \cite{amit, fusistochastic}. We could not find any analytical result on neural networks with non-binary synapses or using the Ising model with plastic interactions in the literature.

Here we present a model of a network of binary point neurons also based on the Ising model. However, in our case we consider the connections between neurons to be plastic so that their strengths change as a result of neural activity. In particular, the fom of the transitions for the coupling constants ressemble a basic Hebbian plasticity rule \cite{gerstnerkistler}. Therefore, it represents a mathematically treatable model capable of reproducing several features from learning and memory in neural networks. 

It is also worth noting that the results presented here can not only be applied to the framework of neural networks but to other models of interacting particles such as systems of interacting spins, models of voters, social networks or infection propagation, among others \cite{liggett}.


The model combines the stochastic dynamics of spins on a finite graph together with the dynamics of the coupling constants between adjacent sites. The dynamics are described by a non-stationary continuous-time Markov process, introduced in Section~\ref{model}. In Section~\ref{results} we prove the transience of the process, and give a sketch of proof of how the transience occurs, i.e. how the system goes to infinity. In the infinite-time dynamics, the absolute values of interaction constants grow to infinite and the spin orientations freeze at a random time each, never to flip again. Finally, Section~\ref{discussion} is regarded to some comments on extensions to the model and predictions on their possible effects.

\section{Model}
\label{model}

Let $G=(V,E)$ be a finite undirected graph without self-loops. For each vertex $v\in V$ we associate a {\it spin} 
$\sigma_v \in \{-1, +1\}$ and, for each edge $e=(v, v') \in E$ we associate a {\it coupling constant} $J_e\equiv J_{vv'} \in \mathbb{Z}$. These constants are often called  exchange energy constants. Here we will also use the term {\it strength} for the coupling constants.

Denote by $\boldsymbol{\sigma}$ a configuration of spins $\bfs = (\sigma_v, v\in V)\in \{-1,1\}^V$ and by $\bf J$ a configuration of strengths $\bfj = (J_e, e\in E)\in \mathbb Z^E$. The state space ${\cal A} $ of the  Markov process  ${\cal M} = (\xi (t), t \in \mathbb R_+)$ is the set of all possible pairs of configurations $(\bfs,\bfj) \in {\cal A} \equiv \{-1,1\}^V\times \mathbb Z^E$. The following functions will play a key role in further definitions
\begin{equation}\label{eta}
\eta_v(\bfs,\bfj)=\sigma_v  \sum_{v': v' \sim v }{J_{vv'} \sigma_{v'} }, \nonumber
\end{equation}
where $v\sim v'$ means that vertices $v$ and $v'$ are neighbors with respect to the graph $G$, i.e. $(v,v') \in E$. We shall omit the arguments $(\bfs,\bfj)$ when it does not lead to confusions.

The transition rates are as follows. Any spin $\sigma_v$ will change its sign $\sigma_v \to -\sigma_v$ with rate
\begin{equation} \label{ci}
c_v(\bfs, \bfj) = \frac{1}{1+\exp{(2 \eta_v)}}.
\end{equation}
Meanwhile, the strength of each edge $(v, v') \in E$, $J_{vv'}$, changes by means of transitions given by
\begin{equation}\label{nu}
J_{vv'} \rightarrow J_{vv'} + \sigma_v \sigma_{v'},
\end{equation}
with constant rate $\nu \in \mathbb{R}_+$ and equal for all edges.
Note that these interaction constants can take both on positive and negative integer values since, for any $(v, v') \in E$, $J_{vv'}$ decreases by one when $\sigma_v=-\sigma_{v'}$ at the transition time and it increases by one when $\sigma_v=\sigma_{v'}$ at that time.

The above transitions define the Markov continuous-time process $\xi(t) = (\bfs(t), \bfj(t))$ on the set space ${\cal A}$. Regarding the initial distribution, spins $\sigma_v$ take values $-1$ or $1$ with the same probability $1/2$, independently, and $\bfj (0) \equiv 0$. 

In order to apply the Lyapunov function technique described below (see \cite{menshikov}) we work with the embedded Markov discrete-time  chain $\hat{\cal M}=(\xi_m, m\in \mathbb N)$ or skeleton of the process ${\cal M}$, with $\xi_m = ( \hat\bfs(m), \hat\bfj(m))$. That alternative description is enough  to prove the transience of the original ${\cal M}$. Instead of the transition rates we define   the corresponding one-step transition probabilities. For any given state $(\bfs,\bfj)$ 
the transition $\sigma_v \to -\sigma_v$ occurs with probability 
$\frac{c_v(\bfs, {\bf J})}{D(\bfs, {\bf J})},$ for any $v \in V$, and 
 the transition $J_{vv'} \rightarrow J_{vv'} + \sigma_{v} \sigma_{v'}$ with probability
$\frac{\nu}{D(\bfs, {\bf J})},$ for any $(v, v') \in E$,
where $D(\bfs, {\bf J})$ is the normalizing constant
\begin{equation}
D(\bfs, {\bf J}) = |E| \nu + \sum_{v \in V}{ c_v( \bfs, {\bf J}) }. \nonumber
\label{D}
\end{equation}
Note that for any state $(\bfs,\bfj)\in \cal A$
\begin{equation}\label{Din}
|E| \nu < D(\bfs, {\bf J}) \le |E| \nu +|V|, \nonumber
\end{equation} 
where $|E|$ and $|V|$ mean the number of edges in $E$ and vertices in $V$, respectively.

\section{Results}
\label{results}
In this section we state and prove the main theorems which regard the transience of the chain $\hat{\cal M}$, and as a consequence, of that of ${\cal M}$, and how this transience occurs, i.e. spins stop changing sign after a finite random time and the absolute value of all the strengths grows to infinity. We also give and prove some necessary lemmas for the proof of the main theorems and include some remarks on the evolution of the chain and extensions to the model.

\begin{theorem}\label{t1}
The Markov chain $\hat{\cal M}$ (${\cal M}$) is transient.
\end{theorem}

In order to prove the transience of the chain $\hat{\cal M}$ we use the Lyapunov function techniques developed in \cite{menshikov}. 
According to Theorem $2.2.2$ from \cite{menshikov} (or Theorem 2.5.8 in \cite{menshikov1}) for a discrete-time Markov chain ${\cal L}=(\zeta_m, m\in \mathbb N)$ with state space $\Sigma$ to be transient it is necessary and sufficient that there exists a measurable positive function (the Lyapunov function) $f(\alpha)$, on the state space, $\alpha \in {\Sigma}$, and a non-empty set $A \subset {\Sigma}$, such that the following inequalities hold
\begin{enumerate}
\item[(L1)]
 $\mathbb{E} \left[ f(\zeta_{m+1}) - f(\zeta_m) | \zeta_m = \alpha \right] \leq 0 \mbox{, } \forall \alpha \notin A,$
\item[(L2)]
$\exists \alpha \notin A : f(\alpha) < \inf_{\beta \in A}{f(\beta)}.$
\end{enumerate}

\noindent
{\bf Proof of Theorem~\ref{t1}.}  Let for $N\in\Z_+$
\begin{equation}\label{fdef}
\displaystyle f_N(\bfs, {\bf J}) = \left\{ \begin{array}{rl} \displaystyle \sum_{v \in V}{\frac{1}{\eta_v}}, & \mbox{ if } \eta_v > N \mbox{, } \forall v \in V \\ \displaystyle \frac{|V|}{N}, & \mbox{ otherwise.} \end{array} \right.
\end{equation}
We show here that there exists $N \in \mathbb{Z}_+$ for which this function and the set $A=A_N$ defined by
\begin{equation}
\label{Adef}
A_N = \{ (\bfs, {\bf J}) \in {\cal A}:  \min_{v\in V} \eta_{v} (\bfs, {\bf J}) \le N \}
\end{equation}
satisfy the above conditions (L1) and (L2).
Theorem~\ref{t1} follows from the following lemma.
\begin{lemma}
Let integer $N_0 >0$ be such that
\begin{equation}\label{N0}
\frac{e^{2N_0}}{N_0+1}\geq\frac{2|V|}{\nu}.
\end{equation}
Then for any $N>N_0$ function \reff{fdef} fulfills conditions {\rm (L1)} and {\rm (L2)}, with the set $A=A_N$.
\end{lemma}
\proof  Let for some fixed $N\ge 1$
\begin{equation}\label{cN}
{\cal C}_N := \{ (\bfs,\bfj)  \in {\cal A}: \  \eta_v > N, \ \forall v \in V \} \equiv {\cal A} \setminus A_N.
\end{equation} 
Suppose firstly that $(\bfs,\bfj) \in {\cal C}_N$. At the first transition time, the system transits from state $(\bfs,\bfj)$ into $(\bfs',\bfj')$. Now consider the transition involving a change of sign for the spin $v$: $\sigma_v \to -\sigma_v$, i.e. $\bfj'=\bfj$ and $\sigma'_u=\sigma_u$ for all $u\in V$ except a site $v$, where $\sigma'_v=-\sigma_v$. In this case the function $f_N(\cdot)$ changes value from $\sum_{w\in V}\frac1{\eta_w}$ to $\frac{|V|}{N}$, since 
\[\eta_v(\bfs',\bfj')=-\sigma_v\sum_{v':\:v'\sim v}J_{vv'}\sigma_{v'}<-N < 0.\]
Assume now that the system transits into a new state $(\bfs',\bfj')$ such that $\bfs'=\bfs$ and $J'_{uw}=J_{uw}$ for all edges $(u,w)\in E$ except one $(v,v')\in E$, namely 
$J'_{vv'}=J_{vv'}+\sigma_v\sigma_{v'}$. It means that 
\[\eta_v(\bfs',\bfj')=\sigma_v\sum_{u:\:u\sim v,u\ne v'}J_{vu}\sigma_{u}+\sigma_v(J_{vv'}+\sigma_{v}\sigma_{v'})\sigma_{v'}=\eta_v(\bfs,\bfj)+1.\]
Similarly, we have for $v'$,
\[\eta_{v'}(\bfs',\bfj')=\eta_{v'}(\bfs,\bfj)+1.\]
Thus
\begin{eqnarray}
\label{proofB1}
&& \mathbb{E}\left[ f(\xi_{m+1}) - f(\xi_m) | \xi_m = (\bfs, {\bf J}) \right] \nonumber \\
&&{} = \sum_{v\in V} \frac{c_v(\bfs, {\bf J})}{D(\bfs, {\bf J})}  \Bigl( \frac{|V|}{N} - \sum_{v \in V}{\frac{1}{\eta_v}} \Bigr) \nonumber  
+ \sum_{(v,v') \in E} \frac{\nu}{D(\bfs, {\bf J})} \Bigl( \frac{1}{\eta_v+1} - \frac{1}{\eta_v} + \frac{1}{\eta_{v'} +1 } - \frac{1}{\eta_{v'}} \Bigr) \nonumber \\
&& {} \le \sum_{v\in V} \frac{c_v(\bfs, {\bf J})}{D(\bfs, {\bf J})}  \frac{|V|}{N} - \frac{1}{2} \frac{\nu}{D(\bfs, {\bf J})}  \sum_{v \in V} \frac{d(v)}{\eta_v(\eta_v+1)} \nonumber \\ 
&& {} = 
\frac{1}{D(\bfs, {\bf J})}  \sum_{v\in V} \Bigl(\frac{|V|}{N(1+\exp(2\eta_v))} - \frac{\nu d(v)}{2 \eta_v (\eta_v+1)} \Bigr),
\end{eqnarray}
where $d(v)$ is the degree of vertex $v$ in the graph $G$. It is easy to observe that for $N\ge N_0$, with $N_0$ fulfilling condition \reff{N0}, 
 any term in (\ref{proofB1}) will be negative since $\eta_v > N$ for all $v\in V$. 

That proves Theorem~\ref{t1}, noting that condition (L2) also holds for function (\ref{fdef}). $\Box$
\vspace{0.5cm}

\begin{rem}
Note that the transition rates for the sign changes given by (\ref{ci}) were chosen in such a way that the classical Gibbs distribution for the Ising model with fixed (not changing in time) coupling constants $(J_e, e\in E)$ would be reversible and consequently invariant.
\end{rem}

Indeed, the classical Hamiltonian for the Ising model is $$ H(\bfs) = -\sum_{v,v': v\sim v'} J_{vv'} \sigma_v \sigma_{v'}.$$
The corresponding Gibbs measure is defined as $\mathbb P_\bfj (\bfs) = \exp \{ -\beta H(\bfs) \}/ Z_\bfj$, where $\beta$ is the usual inverse temperature and $Z_\bfj$ is a normalizing constant. Then, if the weights $(J_e, e\in E)$ are fixed and $\beta=1$, the rates (\ref{ci}) define a Glauber dynamics with reversible Gibbs measure $\mathbb P_\bfj$. 
 
\begin{rem}
In the above presented model, transitions in any of the coupling strengths $J_{vv'}$ with $(v, v' ) \in E$ increase or decrease it by $1$. An interesting extension to the model is to account for jumps of arbitrary fixed size $B \in \mathbb{R}_+$, i.e. the transition given by equation (\ref{nu}) is substituted by
\begin{equation}
J_{vv'} \rightarrow J_{vv'} + \sigma_v \sigma_{v'} B,
\end{equation} so that a single transition in a component of ${\bf J}$ can yield either a very tiny or a very large change in that quantity.
\end{rem}

In this case, Theorem~\ref{t1} also applies with a proof similar to the one given above.

\vspace{1cm}

Now we formulate a theorem that explains how the transience occurs in the model. We  provide a sketch for its proof without a strong proof. 

\begin{theorem}\label{t2} For the discrete time Markov chain $\hat{\mathcal M}$, let $\tau$ be the last time when a sign of configuration $\bfs$ changes:
$\tau := \max \{ m\ge 1:\  \hat\bfs(m-1)\ne\hat\bfs(m)\}$, assuming $\max{\{\emptyset\}} = 0$. Then $\mathbb{P} (\tau <\infty) = 1$ and, as a consequence of that, we obtain that for any $e\in E$ 
\begin{equation}\label{lln}
\lim_{m\to\infty} \frac{|\hat{J}_e(m)|}{m} = \frac{1}{|E|}\ \ \mbox{a.s.}
\end{equation}
\end{theorem}

 \noindent
{\it Sketch of proof of Theorem~\ref{t2}.} Let us remind the transience criteria, i.e. conditions (L1) and (L2) (see Theorem $2.2.2$ in \cite{menshikov} or Theorem 2.5.8 in \cite{menshikov1}). Let $\tau_A$ be the hitting time for the set $A$ as defined in conditions (L1) and (L2). If $f(\cdot)$ is the Lyapunov function for a Markov chain ${\cal L}=(\zeta_m, m\in \mathbb N)$, then from the proof of Theorem~\ref{t1} one gets the following estimation for the probability of the hitting time to be finite (see Lemma 2.5.10 in \cite{menshikov1}):
\begin{equation}\label{tauA}
P(\tau_A < \infty) < \frac{f(\zeta_0)}{\inf_{x\in A} f(x)}. \nonumber
\end{equation}
This estimation can be directly applied in our case. Let $A_N$ be the subset of $\cal A$ defined by (\ref{Adef}), and let $\tau_{A_N}$ be its hitting time. In the construction of the Lyapunov function (\ref{fdef}) we used the set $A_N$ as the set $A$ from conditions (L1) and (L2). Suppose we started from a state $(\bfs,\bfj)\in {\cal C}_N$, then
\begin{equation*}
\mathbb P(\tau_{A_N} < \infty \mid \xi_0 = (\bfs,\bfj)) < \frac{f(\bfs,\bfj)}{\inf_{x\in A_N} f(x)} = \frac{\sum_{v\in V} \eta_{v}^{-1} (\bfs, {\bf J})}{|V|/N} \le \frac{N}{\min_{v\in V} \eta_{v} (\bfs, {\bf J})} \le \frac{N}{N+1} < 1.
\end{equation*}
It means that for any $(\bfs,\bfj)\in {\cal C}_N$, defined by (\ref{cN}), the probability to never hit the set $A_N$ starting from $(\bfs,\bfj)$ is at least $1/(N+1)$:
$$
\mathbb P(\tau_{A_N} = \infty \mid \xi_0 = (\bfs,\bfj) ) \ge \frac{1}{N+1}.
$$
If $\xi_0 = (\bfs,\bfj)$, then the event $(\tau_{A_N} = \infty)$ is equivalent to the event $(\tau=0)$. It means that, at any time, if the process is in the set ${\cal C}_N$, then with probability at least $1/(N+1)$ no sign will ever change again in the spin configuration. Since for any initial state the system will hit ${\cal C}_N$ in a finite time with probability one, this finishes the proof of the almost sure finiteness of the ``freezing'' time $\tau$.


After the ``freezing'' time $\tau$, only transitions in $\hat\bfj$ occur. For any fixed edge $e$ and any time step, with uniform probability $1/|E|$ the edge $e$ is chosen and $|\hat J_e|$ increases by 1. Thus, after the time $\tau$ one has $|\hat J_e(m+1)| = |\hat J_e(m)| + \delta_m$, where $\delta_m$ are i.i.d. random variables with Bernoulli distribution with success probability $1/|E|$. This, altogether with the strong law of large numbers, grants the almost sure convergence given by (\ref{lln}) .
$\Box$

\begin{rem} Theorem~\ref{t2} also holds for the case of a continuous-time Markov process ${\mathcal M}$, with the only difference that the interaction constants fulfill the asimptotic limit
$$
\lim_{t\to\infty} \frac{|J_{vv'}(t)|}{t} = \nu \ \ \mbox{a.s.}
$$
\end{rem}

Theorem~\ref{t2} ensures that there exists the limit 
\begin{equation}\label{sinf}
\lim_{m\to\infty} \hat\bfs(m) = \lim_{t\to\infty} \bfs(t) := \bfs^\infty:=(\sigma_v^\infty, v\in V). \nonumber
\end{equation}
Now let $\chi$ be the sign function 
$$
\chi (x) = \left\{ \begin{array}{ll} +1, & \mbox{ if }x\ge 0; \\ -1,& \mbox{ if }x< 0. \end{array} \right.
$$
By Theorem~\ref{t2}, for any $e\in E$ the following limits exist
\begin{equation}\label{defj}
\lim_{m\to\infty} \chi \bigl( \hat J_{e}(m) \bigr) = \lim_{t\to\infty} \chi \bigl( J_{e}(t) \bigr) =: j_{e}^\infty. \nonumber
\end{equation}
The following theorem gives the relationship between the limiting spin configuration $\bfs^\infty$ and the limiting signs of the interaction constants ${\bf j}^\infty = (j_{e}^\infty, e\in E)$.
\begin{theorem}\label{t3} $j_{vv'}^\infty = \sigma_v^\infty \sigma_{v'}^\infty$ for any $(v,v')\in E$.
\end{theorem}
\noindent
The theorem is a direct consequence of the freezing of the spin configuration.

\section{Discussion}
\label{discussion}

\noindent
Note that the transition rule for the coupling constants given by equation (\ref{nu}) ressembles a Hebbian-like plasticity rule with no homeostasis nor consolidation. Therefore, in the framework of the proposed model, we have proved how such a rule yields an unstable and unbounded growth of the plastic interactions between the set of neurons. These neurons are represented by spins whose dynamics are described by the stochastic Ising model. As a result of this, the spin configuration or neuron states get eventually frozen and none of them undergoes any further transition after some finite time. The addition of constraints into the interaction constants, representing homeostatic or consolidation processes, will thus be necessary in order to bring stability to the network and yield a more biologically realistic behavior \cite{historySTDP, homeostasisreview, amit}.

Our system also reminds of the theory of attractors. However, here the system evolves towards a single but random attractor, $\bfs^\infty$, where it gets frozen never to be able to leave it.

One possible extension to the model is the addition of an external field that is added to the existing mutual interactions between spins. In that case, the evolution of the system is expected to depend strongly on the initial coupling strengths but also on $\nu$. For initial $J_{vv'}=0$ for all $(v, v') \in E$ and small $\nu$, we expect the spins to align with the external field before the coupling strength can be strongly potentiated. In the frozen final configuration probably all or most of spins are aligned with the external field. On the other hand, for large $\nu$ or initial strong couplings few spin flips will occur and the system will probably freeze with the initial spin orientations. 

The addition of bounds and/or other forms of homeostasis to the coupling strengths will change drastically the behavior of the system. In particular, under strong bounds the state space is finite and the chain becomes recurrent, then we expect the coupling strengths to oscillate between minimum and maximum values with time constants determined by $B$ and $\nu$.

\section*{Acknowledgements}
This work was produced as part of the USP project \emph{Mathematics, computation, language and the brain} and the FAPESP (S\~ao Paulo Research Foundation) project \emph{Research, Innovation and Dissemination Center for Neuromathematics} (grant 2013/07699-0) and is supported by the FAPESP grant (2015/10785-0). EP thanks REFI (grant 14-01-00379) and FAPESP (grant 2015/03452-5) for financial support and NUMEC for hospitality. AY also thanks CNPq (Conselho Nacional de Desenvolvimento Cient\'ifico e Tecnol\^ogico) grant 307110/2013-3.

\end{document}